\documentclass[prl,superscriptaddress,showpacs,preprintnumbers,amsmath,amssymb]{revtex4}
\usepackage{color}

\begin{document}
\title{Maximally Entangled States via Mutual Unbiased Collective Bases}

\author{M. Revzen}
\affiliation {Department of Physics, Technion - Israel Institute of Technology, Haifa
32000, Israel}
\date{\today}
\begin{abstract}
Relative and center of mass coordinates are used to generalize mutually unbiased
bases (MUB) and define mutually unbiased collective bases (MUCB). Maximal entangled
states are given as product states in the collective variables.
\end{abstract}


\maketitle

\section {Introduction}
Entanglement is central to some of the intriguing  counter intuitive aspects of
quantum mechanics, yet, no accepted entanglement measure, applicable for both pure
and mixed states, is available. Nontheless for two particles (systems) of equal
(Hilbert space) dimensionality, d, it is intuitively clear that a pure state,
$\psi(1,2)$, whose Schmidt's decomposition \cite{peres} is given by,
\begin{equation}
\psi(1,2)=\frac{1}{\sqrt d}\Sigma_{n=1}^{d}a_nu_n(1)v_n(2),
\end{equation}
with $|a_n|=1$ and $\langle u_n|u_m\rangle=\langle v_n|v_m\rangle=\delta_{n,m}$, is
a {\it maximally entangled} state (MES). Thus taking the partial trace with respect
to one of the particles (e.g. the one labelled by 1) leaves the the other particle
to be, with equal probability, in any state,
\begin{equation}
\rho_2=\Sigma_n\langle
u_n(1)|\psi(1,2)\rangle\langle\psi(1,2)|u_n(1)\rangle=\frac{\Bbb{I}_2}{d}.
\end{equation}

\noindent Two bases $\mathcal{B}_1,\mathcal{B}_2$ are said to be mutually unbiased
(MUB) when the decomposition of a basis vector of one basis, $\mathcal{B}_1$ in
terms of the other's base vectors contains, with equal amplitude, all the base
vectors of $\mathcal{B}_2$. Our aim is to show that such states have a direct
interconnection: we establish a relation between the so called d dimensional set of
mutually unbiased bases (MUB) \cite{wootters,klimov} and maximally entangled states.
This is done via novel MUB that spans the two d dimensional (two) particles Hilbert
spaces with collective coordinates bases that we term MUCB. Thus the $d^2$
dimensional two particles Hilbert sapce is spanned by two, d dimensional collective
coordinates bases. The two (one for each particle) d+1 MUB that are available for
the particles' Hilbert spaces (we consider d=prime) is replaced ($d\ne2$) by two d+1
dimensional MUB sets that account for the collective degrees of freedom: we have one
set of d+1 MUB for the center of mass and one for the relative coordinates degrees
of freedom. The product of of these modes of states is shown to be, almost always,
maximally entangled states. Before tackling our subject we give, in section II, a
brief review of our approach to MUB for both the continuous
 $(d\rightarrow \infty)$ and the finite dimensional, d, Hilbert space. Then in the next
 section, section III, we consider
the continuum case. In this case the rationale involved in this paper is clearest,
in particular, we introduce here the collective bases formalism and define the MUCB.
In section IV we study the d-dimensional case (with d = prime, $\ne2$) where
advantage is taken of the (algebraic) field theory that is applicable here. (The d=2
case is dealt with somewhat differently.) This section contains the definition and
use of collective coordinates and their relation to entanglement for finite
dimensional Hilbert space. The last section is devoted to some remarks and
concluding statements.

\section{Mutual Unbiased Bases (MUB) - brief review}

We briefly summarize some of the MUB features of the continuous,
$d\rightarrow \infty$,
Hilbert space which will be used later.\\
The complete orthonormal eigenfunctions of the quadrature operator,
\begin{equation}\label{X}
\hat{X}_{\theta}\equiv
cos\theta\;\hat{x}+sin\theta\;\hat{p}=U^{\dagger}(\theta)\hat{x} U(\theta),
\end{equation}
with,
\begin{equation}
U(\theta)=e^{i\theta \hat{a}^{\dagger}\hat{a}};\;\;\hat{a}\equiv \frac{1}{\sqrt2}
(\hat{x}+ i\hat{p}),\;\hat{a}^{\dagger}\equiv \frac{1}{\sqrt2}(\hat{x}- i\hat{p}),
\end{equation}
are labelled by $|x,\theta\rangle$:
\begin{equation}\label{X1}
\hat{X}_{\theta}|x,\theta\rangle=x|x,\theta\rangle.
\end{equation}
(As is well known \cite{ulf} $\hat{a},\;\hat{a}^{\dagger}$ are referred to as
"creation" and annihilation" operators, respectively.) This (Eq.(\ref{X},\ref{X1})
allow us to relate bases of different labels much like time evolution governed by an
harmonic oscillators hamiltonian \cite{larry}: we may view the basis labelled by
$\theta$ as "evolved" from one labelled by $\theta=0$ i.e. for a vector in the
x-representation we have,
\begin{equation}\label{evol}
|x,\theta \rangle= U^{\dagger}(\theta)|x\rangle.
\end{equation}
We note that the state whose eigenvalue is x  in a basis labelled by $\theta$, i.e.
$|x,\theta\rangle$, are related to the eigenfunction, $|p\rangle,$ of the momentum
operator via,
\begin{equation}\label{pi/2}
|x,\frac{\pi}{2}\rangle=|p\rangle.
\end{equation}
(In this equation $|p\rangle$ is the eigenfunction of the momentum operator,
$\hat{p}$, whose eigenvalue ,p, is numerically equal to x.) Similarly we have,
\begin{equation}\label{reverse}
|x,\pm \pi\rangle=|-x\rangle,
\end{equation}
i.e. evolution by $\pm \pi$ may be viewed as leading to a vector in the same basis
(i.e. $\theta$ intact) but evolves to a vector whose eigenvalue is of opposite sign.
Returning to Eq.(\ref{evol}) we utilize the known analysis of the evolution operator
$U(\theta)$ \cite{larry} to deduce that, in terms of the eigenfunction of
$\hat{Y}_0=\hat{x}$, viz the x-representation, $|y;\theta\rangle$ is given by,

\begin{equation} \label{ls}
\langle x|y,\theta \rangle=\langle
x|U^{\dagger}(\theta)|y\rangle=\frac{1}{\sqrt{2\pi sin \theta}}
e^{-\frac{i}{2sin\theta}\left([x^2+y^2]cos \theta-2xy\right)}.
\end{equation}
These states form a set of MUB each labelled by $\theta$:
\begin{equation}\label{mub}
|\langle x;\theta|y, \theta' \rangle|=\frac{1}{\sqrt{2\pi |sin(\theta-\theta'|}}.
\end{equation}
Thus the verification of the particle as being in the state of coordinate x in the
basis labelled by $\theta$ implies that it is equally likely to be in any coordinate
state x' in the basis labelled by $\theta'$ ($\theta'\ne \theta)$. Note however that
in the continuum $(d \rightarrow \infty)$ considered above, the inter-basis scalar
product, Eq.(\ref{mub}), retains, in general, their basis labels ($\theta,\theta')$.
For a finite, d dimensional Hilbert space the scalar inter MUB product is, in
absolute value, $\frac{1}{\sqrt d}$ and does not contain any information on the base
labels \cite{amir}. It was shown by Schwinger \cite{schwinger} that complete
operator basis (COB) for this problem constitute of $\hat{Z}$ and $\hat{X}$ with,
\begin{equation}
\hat{Z}|n\rangle=\omega^n |n\rangle,\;\;\omega=e^{i\frac{2\pi}{d}},\;\;\hat{X}|n\rangle=
|n+1\rangle\;\;|n+d\rangle =|n\rangle.
\end{equation}
It was further shown \cite{ivanovich,wootters,tal,klimov} that the maximal number of
MUB possible for a d dimensional Hilbert space is d+1. However only for d=prime (or
a power of a prime) d+1 such bases are known to exist. (3 such bases are known for
all values of d.) For the case of d=prime a general MUB basis is given in terms of
the computational basis,\cite{tal}
\begin{equation}
|m;b\rangle=\frac{1}{\sqrt d}\Sigma_0^{d-1}\omega^{\frac{b}{2}n(n-1)-nm}|n\rangle.
\end{equation}
These are the eigenfunction of $\hat{X}\hat{Z}^{b}$, $b=0,1,...d-1$. Here b may be
used to label the basis. (These d bases supplemented with the computational basis
form the d+1 MUB , \cite{tal}.)

\section{The continuum, $d\rightarrow \infty$, case}

The generic maximally entangled state is the EPR \cite{epr} state,
\begin{equation}
|\xi,\mu \rangle=\frac{1}{\sqrt{2\pi}}\int dx{_1}dx{_2}\delta\big(\frac{x_1-x_2}
{\sqrt2}-\xi\big)e^{i\mu\frac{x_1+x_2}{\sqrt2}}|x_1\rangle |x_2\rangle,
\end{equation}
($\sqrt2$ is introduced for later convenience.) We now consider an alternative means
of accounting for the two particles states to which we refer to as the "relative"
and "center of mass" coordinates (we assume equal masses for simplicity),
\begin{equation}
\xi=\frac{x{_1}-x{_2}}{\sqrt2};\;\;\eta=\frac{x_1+x_2}{\sqrt2}.
\end{equation}
The corresponding operators, each acting on one of these coordinates, are
\begin{equation}\label{xieta}
\hat{\xi}=\frac{\hat{x}_1-\hat{x}_2}{\sqrt2};\;\;\eta=\frac{\hat{x}_1+\hat{x}_2}{\sqrt2},
\end{equation}
with,
\begin{equation}
\hat{\xi}|\xi\rangle=\xi|\xi\rangle;\;\;\hat{\eta}|\eta\rangle=\eta|\eta\rangle.
\end{equation}
Using relations of the type,
\begin{equation}
\langle x_1x_2|\hat{\xi}|\xi\eta\rangle=\xi\langle x_1x_2|\xi\eta\rangle=
\langle x_1x_2|\frac{\hat{x}_1-\hat{x}_2}{\sqrt2}|\xi\eta\rangle=
\frac{x_1-x_2}{\sqrt2}\langle x_1x_2|\xi\eta\rangle,
\end{equation}
One may show that,
\begin{equation}
\langle x_1x_2|\xi\eta\rangle=\delta\big(\xi-\frac{x_1-x_2}{\sqrt2}\big)\delta\big
(\eta-\frac{x_1+x_2}{\sqrt2}\big).
\end{equation}
We note that $\hat{x}_1,\hat{p}_1$ form a complete operator basis (COB) for the
first particle Hilbert space (we do not involve spin) and similarly
$\hat{x}_2,\hat{p}_2$ for the second particle, i.e.,
\begin{eqnarray}
\left[\hat{x}_1,\hat{p}_1\right] &=& \left[\hat{x}_2,\hat{p}_2\right]=i,\;\;\nonumber \\
 \left[\hat{x}_1,\hat{p}_2\right] &=& \left[\hat{x}_2,\hat{p}_1\right]=
 \left[\hat{x}_2,\hat{x}_1\right]=\left[\hat{p}_2,\hat   {p}_1\right]=0,
\end{eqnarray}
thus we have that the two pairs of operators form a COB for the
combined ($d^2$ dimensional) Hilbert space. Defining,
\begin{equation}\label{numu}
\hat{\nu}\equiv\frac{\hat{p}_1-\hat{p}_2}{\sqrt2},\;\;\hat{\mu}\equiv\frac{\hat{p}_1+
\hat{p}_2}{\sqrt2},
\end{equation}
we have
\begin{eqnarray}\label{comm}
\left[\hat{\xi},\hat{\nu}\right]&=&\left[\hat{\eta},\hat{\mu}\right]=i,\;\;\nonumber \\
\left[\hat{\xi},\hat{\mu}\right]&=&\left[\hat{\xi},\hat{\eta}\right]=
\left[\hat{\eta},\hat{\nu}\right]=\left[\hat{\mu},\hat{\nu}\right]=0.
\end{eqnarray}
These (viz: $\hat{\xi},\hat{\nu},\hat{\eta},\hat{\mu}$) form an
alternative COB for the (combined) Hilbert space with
$\hat{\xi},\hat{\nu}$ spanning the relative coordinates space
while $\hat{\eta},\hat{\mu}$ the "center of mass" one. By analogy
with the single particle state analysis we now define "creation"
and "annihilation" operators for the collective degrees of
freedom:
\begin{eqnarray}
\hat{A}&=&\frac{1}{\sqrt2}(\hat{\xi}+i\hat{\nu}),\;\;\hat{A}^{\dagger}=
\frac{1}{\sqrt2}(\hat{\xi}-i\hat{\nu}),\nonumber\\
\hat{B}&=&\frac{1}{\sqrt2}(\hat{\eta}+i\hat{\mu}),\;\;\hat{B}^{\dagger}=
\frac{1}{\sqrt2}(\hat{\eta}-i\hat{\mu}),
\end{eqnarray}
these abide by the commutation relations
\begin{equation}
\left[\hat{A},\hat{A}^{\dagger}\right]=\left[\hat{B},\hat{B}^{\dagger}\right]=1,
\end{equation}
with all other commutators vanishing, and the "evolution"
(Eq.(\ref{evol})) operators are,
\begin{equation}
V_A(\theta)=e^{i\theta
\hat{A}^{\dagger}\hat{A}};\;\;V_B(\theta)=e^{i\theta
\hat{B}^{\dagger}\hat{B}}.
\end{equation}
These operators are (as we shall see shortly) our entangling operators: each (pair)
act on different "collective" coordinate. We note that for $\theta=\theta'$ and only
in this case,
\begin{equation}\label{product}
V_A^{\dagger}(\theta)V_B^{\dagger}(\theta)=U_1^{\dagger}(\theta)U_2^{\dagger}(\theta),
\end{equation}
i.e. in this case a simple relation exists between the particles' operators and the
collective ones. The results of section II, Eq.(\ref{evol}), now read,
\begin{eqnarray}
|\xi,\theta\rangle&=& V_A^{\dagger}(\theta)|\xi\rangle, \nonumber\\
|\eta,\theta'\rangle&=& V_B^{\dagger}(\theta')|\eta\rangle.
\end{eqnarray}
The commutation relation, Eq.(\ref{comm}), implies  that the basis
$|\eta\rangle,$ the eigenbasis of $\hat{\eta}$ (i.e.
$V_B^{\dagger}(0)|\eta\rangle$), and the basis $|\mu\rangle,$ the
eigenstates of $\hat{\mu},$ (i.e. the states
$V_B^{\dagger}(\frac{\pi}{2}))|\eta\rangle$, are MUB with,
\begin{equation}
\langle\eta|\mu\rangle=\frac{1}{\sqrt{2\pi}}e^{i\eta\mu},
\end{equation}
With similar expression for $\langle \xi|\nu\rangle$. Note that, in our approach,
 these follow from the equations that corresponds to Eq.(\ref{ls}). We have then that the
 maximally entangled state (the EPR state)
$$|\xi\rangle|\mu\rangle$$
is a product state in the collective variables. It is natural now to consider mutual
unbiased collective bases (MUCB) labelled, likewise, with $\theta$: The relative
coordinates bases $V_A^{\dagger}(\theta)|\xi\rangle$ is one such MUCB. The center of
mass $V_B^{\dagger}(\theta)|\eta\rangle$ is another. We now formulate our link
between MUB and (maximal) entanglement thus consider the (product) two particle
state $|x_1\rangle |x_2\rangle$. It may be written in terms of a product state in
the "collective" coordinates: (When clarity requires we shall mark henceforth the
eigenstates of the collective operators with double angular signs,
$\rangle\rangle,$.)
\begin{equation}
|x_1\rangle |x_2\rangle=\int d\xi'd\eta'\langle \xi',\eta'|x_1,x_2\rangle
|\xi'\rangle\eta'\rangle\;=\;|\xi=\frac{x_1-x_2}{\sqrt2}\rangle\rangle|\eta=
\frac{x_1+x_2}{\sqrt 2}\rangle\rangle.
\end{equation}

We now assert that replacing the basis $|\eta\rangle$ by any of the MUB bases,
$$|\eta\rangle\rightarrow|\eta,\theta\rangle =
V_B^{\dagger}(\theta)|\eta\rangle,\;\theta\ne0,$$
give a maximally entangled state: $|\xi\rangle|\eta,\theta\rangle$.  (The EPR state,
$|\xi,\mu\rangle$ is the special case of $V_B^{\dagger}(\frac{\pi}{2})$.)   The proof
is most informative  with the state
$|\xi\rangle|\mu,\theta\rangle$: (Note: $|\mu,\theta\rangle=V_{B}(\theta)|\mu\rangle=
V_{B}(\theta+\frac{\pi}{2})|\eta\rangle$.)
\begin{eqnarray} \label{maxentg}
|\xi\rangle|\mu,\theta\rangle&=&\int dx_1dx_2|x_1,x_2\rangle\langle
x_1,x_2|\int d\eta d\bar{\eta}\;|\xi,\eta\rangle\langle\eta|\bar{\eta},\theta\rangle
d\bar{\eta}\langle\bar{\eta},\theta|\mu,\theta\rangle \nonumber \\
    &=&\frac{\sqrt2}{2\pi\cos\theta}e^{\frac{i\mu}{2\cos\theta}\big(2\xi-\mu \sin\theta\big)}
    \int dxe^{\frac{\sqrt2 i x\mu}{cos\theta}}
    |x\rangle|x-\sqrt2 \xi\rangle.
\end{eqnarray}
The various matrix elements are given by,
\begin{eqnarray}
\langle x_1|\xi,\eta \rangle&=&\delta\left( x_1-\frac{\eta+\xi}{\sqrt2}\right)
\Big|x_2=\frac{\eta-\xi}{\sqrt2}\Big\rangle,\nonumber \\
\langle \eta|\bar{\eta},\theta \rangle&=&\frac{1}{\sqrt{2\pi|sin\theta|}}
e^{-\frac{i}{2sin\theta}\big[(\eta^2+\bar{\eta}^2)cos\theta -2\bar{\eta}\eta \big]},
\nonumber \\
\langle\bar{\eta},\theta|\mu,\theta\rangle&=&\frac{1}{\sqrt{2\pi}}e^{i\bar{\eta}\mu}.
\end{eqnarray}
The state, Eq.(\ref{maxentg}), upon proper normalization, is the maximally entangled
EPR state, as claimed (cf. Appendix B): it involves, with equal probability, all the
vectors of the x representation. It follows by inspection that this remain valid to
all states (exceptions are specific angles that are specified below) build with
MUCB,
\begin{equation}
|\xi,\theta\rangle|\eta,\theta'\rangle.
\end{equation}
We summarize our consideration thus far as follows: Consider two pairs of operators
(we assume that these two form a COB) pertaining to two Hilbert spaces. Each pair is
made up of {\it non commuting} operators, e.g. $\hat{x}_1,\hat{p}_1$ and
$\hat{x}_2,\hat{p}_2$. Now form two {\it commuting} pairs of operators with these
operators as their constituents, e.g. $\hat{R}_A(0)=\hat{x}_1-\hat{x}_2$ and
$\hat{R}_B(\frac{\pi}{2})= \hat{p}_1+\hat{p}_2$: the common eigenfunction of
$\hat{R}_A(0)$ and $\hat{R}_B(\frac{\pi}{2})$ is, necessarily, an entangled state.
This was generalized via the consideration of the common eigenfunction of the
commuting operators
\begin{eqnarray}
R_A(\theta)&\equiv& V^{\dagger}_{A}(\theta)\hat{\xi}V_{A}(\theta)=
cos\theta \hat{\xi}+sin\theta \hat{\nu}, \nonumber \\
R_B(\theta') &\equiv&V^{\dagger}_B(\theta')\hat{\eta}V_B(\theta')=
cos\theta' \hat{\eta}+sin\theta' \hat{\mu}.
\end{eqnarray}
These commute for all $\theta, \theta'$ and thus have common eigenfunctions. For
$\theta=\theta'\;and\;\theta=\theta'\pm \pi$ and only for these values, the common
eigenfuction is a product state (in these cases the constituents commute, e.g. $
\hat{x}_1-\hat{x}_2$ and $\hat{x}_1+\hat{x}_2$). This is shown in Appendix C. For
all other $\theta, \theta'$ the common eigenfunction is an entangled state. (
Moreover, these states are maximally entangled states. The proof is outlined in
Appendix B.) The definition of the "collective" coordinates is such as to assure the
decoupling of the  combined Hilbert space to two independent subspaces whose
constituent (pairs) operators commute (e.g. $ \hat{x}_1-\hat{x}_2$ and
$\hat{x}_1+\hat{x}_2$) much as it (the Hilbert space) was decoupled with the
individual particles operators.

 \section{Finite dimensional analysis - collective coordinates}

We now turn to the more intriguing cases of d dimensional Hilbert spaces.
We confine our study to (two) d-dimensional spaces with d a prime ($\ne2$). The indices
are elements of an algebraic field of order d. The computational, two particle, basis
states  $$|n\rangle_1 |m\rangle_2\;\;n,m=0,1,..d-1.$$ spans the space. A COB (complete
operator basis) is defined via ($i=1,2$,
\begin{eqnarray}
Z_i|n\rangle_{i}&=&\omega^{n_i}|n\rangle_{i},\;\;\omega=e^{i\frac{2\pi}{d}} \nonumber \\
X_i|n\rangle_{i}&=&|n+1\rangle_{i},
\end{eqnarray}

We now define our collective coordinate operators via,
\begin{equation}\label{coll}
\bar{Z}_1\equiv Z_1^{\frac{1}{2}}Z_2^{-\frac{1}{2}};\;\bar{Z}_2\equiv Z_1^{\frac{1}{2}}
Z_2^{\frac{1}{2}}.
\end{equation}
(We remind the reader that the exponent value of $\frac{1}{2}$ is a field number such
that twice its value is  1 mode[d], e.g. for d=7,  $\frac{1}{2}=4.$)
Eq.(\ref{coll}) implies that,
\begin{equation}
Z_1=\bar{Z}_1\bar{Z}_2,\;\;Z_2=\bar{Z}_1^{-1}\bar{Z}_2.
\end{equation}
The spectrum of $\bar{Z}_i$ is $\omega^{\bar{n}}, \bar{n}=0,1,..d-1$ since we have that
$\bar{Z}_i^{d}=1$
 and we consider the bases that diagonalize $\bar{Z}_i$:
\begin{equation}
\bar{Z}_i|\bar{n}_i\rangle=\omega^{\bar{n}_i}|\bar{n}_i\rangle.
\end{equation}
To obtain the transformation function $\langle n_1, n_2|\bar{n}_1,\bar{n}_2\rangle$
we evaluate $\langle n_1,n_2|A|\bar{n}_1,\bar{n}_2\rangle$  with A equals
$Z_1,Z_2,\bar{Z}_1,\bar{Z}_2$ in succession. e.g. for $A=Z_1$,
\begin{equation}
\langle n_1, n_2|Z_1|\bar{n}_1,\bar{n}_2\rangle=\omega^{n_1}\langle n_1,n_2|\bar{n}_1,
\bar{n}_2\rangle=\omega^{\bar{n}_1+\bar{n}_2}\langle n_1,n_2|\bar{n}_1,\bar{n}_2\rangle.
\end{equation}
These give us the following relations (all equations are modular: mode[d]),
\begin{eqnarray}
n_1&=&\bar{n}_1+\bar{n}_2;\;\;n_2=-\bar{n}_1+\bar{n}_2, \nonumber \\
\bar{n}_1&=&\frac{n_1}{2}-\frac{n_2}{2};\;\;\bar{n}_2=\frac{n_1}{2}+\frac{n_2}{2}.
\end{eqnarray}
Whence we deduce,
\begin{equation}
\langle n_1,n_2|\bar{n}_1,\bar{n}_2\rangle=\delta_{n_1,\bar{n}_1+\bar{n}_2}
\delta_{n_2,\bar{n}_1-\bar{n}_2}.
\end{equation}
In a similar fashion we now define,
\begin{equation}
\bar{X}_1\equiv X_1X_2^{-1},\;\;\bar{X}_2\equiv X_1X_2\;\rightarrow X_1=
\bar{X}_1^{1/2}\bar{X}_2^{1/2},\;X_2=\bar{X}_1^{-1/2}\bar{X}_2^{1/2}.
\end{equation}
These entail,
\begin{eqnarray}
\bar{X}_i\bar{Z}_i&=&\omega\bar{Z}_i\bar{X}_i,\;i=1,2 \nonumber \\
\bar{X}_i\bar{Z}_j&=&\bar{Z}_j\bar{X}_i ,\;i \ne j.
\end{eqnarray}
Thence,
\begin{equation}
\bar{X}_i|\bar{n}_i\rangle=|\bar{n}_i+1\rangle,\;\;i=1,2
\end{equation}
and, denoting the eigenvectors of the barred operators (i.e. the collective coordinates)
with double angular sign we have that
\begin{eqnarray}
\bar{X}_1|n_1,n_2\rangle&=&\bar{X}_1|\frac{n_1-n_2}{2},\frac{n_1+n_2}{2}\rangle\rangle=
|\frac{n_1-n_2}{2}+1,\frac{n_1+n_2}{2}\rangle\rangle \nonumber \\
\bar{X}_2|n_1,n_2\rangle&=&\bar{X}_2|\frac{n_1-n_2}{2},\frac{n_1+n_2}{2}\rangle\rangle=
|\frac{n_1-n_2}{2},\frac{n_1+n_2}{2}+1\rangle\rangle .
\end{eqnarray}
Recalling, Eq.(10), the set of MUB associated with
$|\bar{n}_2\rangle\rangle$, viz $|\bar{n}_2,b\rangle\rangle$ (with
$b=0,1..d-1$):
\begin{equation}
|\bar{n}_2,b\rangle=\frac{1}{\sqrt
d}\Sigma_{\bar{n}}\omega^{\frac{b}{2}\bar{n}(\bar{n}+1)
-\bar{n}\bar{n}_2}|\bar{n}\rangle\rangle.
\end{equation}
This state is an eigenfunction of $\bar{X}_2\bar{Z}_2^b$, cf Eq. ().
Our association of maximally entangled states with MUB amounts to the following. Given a
product state. We write it as a product state of the collective coordinates, e.g.
\begin{equation}
|n_1\rangle|n_2\rangle=|\bar{n}_1\rangle\rangle|\bar{n}_2\rangle\rangle,\;n_1=
\bar{n}_1+\bar{n}_2;\;n_2=\bar{n}_2-\bar{n}_1.
\end{equation}
Now replace one of these (collective coordinates states) by a state (any one of which)
belonging to its MUB set, e.g.
\begin{equation}
|\bar{n}_1\rangle\rangle|\bar{n}_2\rangle\rangle \rightarrow|\bar{n}_1\rangle\rangle
|\bar{n}_2,b\rangle\rangle,\;\;b=1,2..d-1.
\end{equation}
The resultant state is a maximally entangled state. We prove it for a representative
example by showing that measuring in such state $Z_1$ that yield the value $n_1$ leaves
the state an eigenstate of $Z_2$ with a specific eigenvalue. To this end we consider the
projection of the state $\langle n_1|$ on the representative state . Somewhat lengthy
calculation yields,
\begin{equation}
\langle
n_1|\bar{n}_1\rangle\rangle|\bar{n}_2,b\rangle\rangle=\frac{1}{\sqrt
d}|n_2=-2\bar{n}_1+n_1\rangle \omega^{\frac{b}{2}(n_1-\bar{n}_1)(
n_1-\bar{n}_1-1)-\bar{n}_2(n_1-\bar{n}_1)}.
\end{equation}
Here the state $|n_2=-2\bar{n}_1+n_1\rangle$ is an eigenstate of $Z_2$ proving our point.\\

We discuss now the finite dimensional Hilbert space in a manner that stresses its analogy
with the
 $d\rightarrow \infty$ case considered above: Given two, each d-dimensional, Hilbert spaces
 and each
 pertaining to one of two particles (systems) bases. The combined, $d^2$-dimensional space
 is conveniently spanned by a basis made of product of computational bases,
 $|n_1\rangle|n_2\rangle;\;n_i=0,1,...d-1$. Each of the computational basis may be
 replaced by any of the d other available MUB bases (recall that we limit ourselves to
 d=prime where d+1 MUB are available \cite{tal}). Each MUB basis is associated \cite{tal}
 with a unitary operator, $X_iZ_i^b,\;\;b=0,1,..d-1$ (these supplemented by $Z_i$ account
 for the d+1 MUB). We have shown above that the combined Hilbert space may be accounted
 for by what we termed collective coordinates computational bases:
 $|\bar{n}_1\rangle| \bar{n}_2\rangle,\;\;\bar{n}_i=0,1,...d-1.$ (Here  $|\bar{n}_1\rangle$
 relates to the "relative" while $|\bar{n}_2\rangle$ to the "center of mass" coordinate.)
 These were defined such that  $$|n_1\rangle|n_2\rangle=
 |\bar{n}_1\rangle\rangle| \bar{n}_2\rangle\rangle.$$ We then noted that, in analogy
 with the $R_i(\theta)$ of the $d\rightarrow \infty$ case each $|\bar{n}_i\rangle\rangle$
 may be replaced by any of the d+1 MUB of the collective coordinates. These are
 associated with $\bar{X}_i\bar{Z}_i^b,\;\;b=0,1,...d-1$. We now have the space spanned
 by $|\bar{n}_1,b_1\rangle\rangle|\bar{n}_2,b_2\rangle\rangle$. These except for
 "isolated" combination are maximally entangled states (cf. Appendix A). The isolated
 values are the $b_1=b_2$ cases and the bases associated with
 $\bar{X}_1\bar{Z}_1^b\;\;and\;\;\bar{X}_2^{-1}\bar{Z}_2^{-b}$ - the eigenstates of
 which are product states.\\
Now while in the finite dimensional case the set of d+1 MUB states can be
constructed only for d a prime
 (or a power of a prime - which is not studied here) no such limit holds for the continuous case. The intriguing
 price being that in this ($d\rightarrow \infty$) case the definition of the MUB states involves a basis dependent
 normalization. We have considered the cases with d=prime. The case d=2 need special treatment because, in this
 case, +1=-1 [mode 2] (indeed 2=0 [mode 2]) hence the "center of mass" and "relative" coordinates are indistinguishable.
 Here the operator vantage point may be used to interpret the known results \cite{schwinger}. The
 operators vantage point involves the following: given two systems $\alpha,\beta$. Consider two non commuting
 operators pertaining to $\alpha: \;A,A'$ and correspondingly two non-commuting operators B and B' that belong to
 $\beta$. our scheme was to construct a common eigenfunction for AB and A'B' with (which we assume) $[AB,A'B']=0$.
 This common eigenfunctions are maximally entangled. This is trivially accomplished: e.g. consider ($\alpha,\beta\rightarrow1,2$
 $\sigma_{x1}\sigma_{x2}\;\;with\; \sigma_{z1}\sigma_{z2}$, and  $\sigma_{x1}\sigma_{x2}\;\;with\;\sigma_{1}\sigma_{y2}$.
 Their common eigenfunctions are the well known Bell states \cite{sam}.





\section{Concluding Remarks}

An association of maximally entangled states for two particles, each of
dimensionality d, with mutually unbiased bases (MUB) of d dimensional Hilbert space
inclusive of the continuous ($d\rightarrow \infty$) cases were established. The
analysis is based on the alternative forms for the two particle state: product of
computational based states, and a product of the state given in terms of collective
coordinates (dubbed center of mass and relative). A formalism allowing such an
alternative accounting for the states was developed for d a prime ($\ne 2$)  which
applies the finite, d ($\ne2$), dimensional cases where the maximally allowed MUB
(d+1) is known to be available. Based on the alternative ways of writing the two
particle states we defined and demonstrated the use of mutually unbiased collective
bases (MUCB). The latter is generated  by noting that replacing one of the states in
the collective coordinates product state with any of its MUCB states realizes a
maximal entangled state. Such state is, by construction, made of eigenfunctions of
commuting pairs of two particles operators with the single particle operators in the
different pair non commuting. Thus we shown that maximally entangled states both in
the continuum and some finite dimension Hilbert spaces may be viewed as product
states in collective variables and have demonstrated the intimate connection between
entanglement and operator non commutativity (i.e. the uncertainty principle).

\section* {Appendix A: Maximally Entangled State}
We prove here that the state $|\xi\rangle|\eta,\frac{\pi}{2}\rangle$ is a maximally
entangled state. (Note $|\eta,\theta+\frac{\pi}{2}\rangle =|\mu,\theta\rangle$).
This can be seen directly by calculating the $x$ representation of the state and
noting that it is of the same form of the EPR state, i.e. its Schmidt decomposition
contains all the states paired with coefficients of equal magnitude
\cite{peres,shim}:
\begin{eqnarray}
|\xi\rangle|\mu,\theta\rangle&=&\int dx_1dx_2|x_1,x_2\rangle \langle
x_1,x_2|\int d\eta d\bar{\eta}\;|\xi,\eta\rangle\langle\eta|\bar{\eta},\theta\rangle
d\bar{\eta}\langle\bar{\eta},\theta|\mu,\theta\rangle \nonumber \\
    &=&\frac{\sqrt2}{2\pi\cos\theta}e^{\frac{i\mu}{2\cos
    \theta}\big(2\xi-\mu \sin\theta\big)}
    \int dxe^{\frac{\sqrt2 i x\mu}{cos\theta}}
    |x\rangle|x-\sqrt2 \xi\rangle.
\end{eqnarray}
This is a maximally entangled state for $0\le\theta<\frac{\pi}{2}$.  Now considering
the state for $\theta=\frac{\pi}{2}$ we have (c.f., Eq.(\ref{pi/2},\ref{reverse})
\begin{equation}
\langle x_1,x_2|\xi\rangle|\mu,\frac{\pi}{2}\rangle=\langle
x_1,x_2|\xi,-\eta\rangle=\delta\Big(x_1-\frac{\xi-\eta}{\sqrt2}\Big)\Big(x_2+\frac{\xi+
\eta}{\sqrt2}\Big),
\end{equation}
i.e. at $\theta=\frac{\pi}{2}$  the state is a product state. We interpret this to
mean that entanglement is not analytic \cite{amir}.\\

\section*{Appendix B: Maximal entanglement of the state $|\xi,\theta;\eta,\theta'\rangle$}

We now prove that the state $|\xi,\theta,\eta,\theta'\rangle$ is a maximally entangled
state for all
 $\theta,\theta'$ (except for isolated points:$\theta=\theta'\pm \pi$, at these points
 the state
is a product state). We note that
\begin{equation}\label{quad}
\hat{A}^{\dagger}\hat{A}+\hat{B}^{\dagger}\hat{B}=\hat{a}^{\dagger}_1\hat{a}_1+
\hat{a}^{\dagger}_2\hat{a}_2.
\end{equation}
Hence, cf. Eq. (\ref{product}), here the numerical subscripts refers to the
particles,
\begin{equation}
V_A^{\dagger}(\theta)V_B^{\dagger}(\theta)=U_1^{\dagger}(\theta)U_2^{\dagger}(\theta).
\end{equation}
Assuming, without loss of generality that $\theta' > \theta$ (when they are equal the
state is a product state), we may thus write ($\Delta=\theta'-\theta$),
\begin{equation}\label{maxent}
|\xi,\theta;\eta,\theta'\rangle=\int d\bar{\eta}\Big|x_1=\frac{\bar{\eta}+\xi}{\sqrt2},
\theta\big\rangle \Big|x_2=\frac{\bar{\eta}-\xi}{\sqrt2},\theta\big\rangle
\frac{1}{\sqrt{2\pi|sin\Delta|}}e^{-\frac{i}{2sin\Delta}\big[(\eta^2+\bar{\eta}^2)
cos\Delta -2\bar{\eta}\eta \big]}.
\end{equation}
Here the vectors ($|x_i\rangle$) are the single particle eigenvectors of
$U^{\dagger}(\theta)\hat{x}_iU(\theta)$.
Now our proof that the state $|\xi,\theta;\eta,\theta'\rangle$ is a maximally entangled
state is attained via  "measuring" the position of the first particle (in the basis
labelled by $\theta$), i.e. calculating the projection $\langle x'_1,\theta|\xi,\theta;
\eta,\theta'\rangle$, and showing that the resultant state is the second particle in a
definite (up to a phase factor) one particle state, $|y_2,\bar{\theta}\rangle$ with $y_2$
linearly related to $x'_1$. Thus ($x'=x'_1$):
\begin{equation}
\langle x'_1|\xi,\theta;\eta,\theta'\rangle= \frac{1}{\sqrt{\pi|sin\Delta|}}
e^{\frac{i}{2sin\Delta}\big[2\xi\eta+(\xi^2+\eta^2)cos\Delta\big]}
e^{\frac{i}{sin\Delta}\big[(x'^2-\sqrt2 x'\xi)cos\Delta-\sqrt2 x'\eta\big]}|x'-
\sqrt2 \xi\rangle.
\end{equation}
QED\\
\section*{Appendix C: Angular labels for product states}
The proof that $|\xi,\theta\rangle|\eta,\theta'\rangle$ are
product states for $\theta=\theta'\;and\;\theta=\theta'\pm \pi$
utilizes the following preliminary observations:\\
a. $V_A^{\dagger}(\pm \pi)|\xi \rangle=|-\xi\rangle;\;V_B^{\dagger}(\pm
\pi)|\eta\rangle=|-\eta\rangle$ i.e. "evolution" by $\pm \pi$ may be viewed as
leaving the basis unchanged but "evolves" to a state whose eigenvalue is of opposite
sign. See Eq. (\ref{reverse}).\\
The states $|\xi \rangle|\pm \eta\rangle,\;|\pm \xi\rangle|\eta \rangle$ are product
states: e.g.
$$|\xi\rangle|-\eta\rangle=\int dx_1dx_2|x_1\rangle_1|x_2\rangle\langle
x_1|\langle x|\xi\rangle|-\eta\rangle=$$
$$\int
dx_1dx_2|x_1\rangle|x_2\rangle\delta\big(\xi-\frac{x_1-x_2}{\sqrt2}\big)\delta
\big(-\eta-\frac{x_1+x_2}{\sqrt2}\big)=\big|\frac{\xi-\eta}{\sqrt2}\big\rangle\big|-\frac{\eta+\xi}{\sqrt
2}\big\rangle.$$ QED.\\
These observations imply that, e.g.,
$$|\xi,\theta\rangle|\eta,\theta+\pi\rangle=V_A^{\dagger}(\theta)V_B^{\dagger}(\theta+\pi)|\xi\rangle|\eta\rangle=$$
$$U_1^{\dagger}(\theta)U_2^{\dagger}|\xi\rangle|-\eta\rangle=\big|\frac{\xi-\eta}{\sqrt2};\theta\big\rangle
\big|-\frac{\eta+\xi}{\sqrt2};\theta \big\rangle.$$ With similar results for $|\pm \xi\rangle
|\eta\rangle$. These are are product states each involves a distinct particle.\\

Acknowledgments: Informative comments by O. Kenneth and C. Bennett
are gratefully acknowledged.


\begin{thebibliography}{99}

\bibitem{peres} A. Peres, {\it Quantum Theory: Concepts and Methods} (Kluwer, Dordrect,
1995).
\bibitem{wootters} W. K. Wootters, Ann. Phys.,(N.Y.) {\bf 176}, 1 (1987),  W. K. Wootters and  Fields, Ann. Phys.,(N.Y.), K. S. Gibbons, M. J. Hoffman and W. K. Wootters, Phys. Rev. A {\bf 70}, 062101 (2004).

\bibitem{klimov} A. B. Klimov, L. L. Sanchez-Soto and H. de Guise, J. Phys. A: Math.
Gen. {\bf 38}, 2747 (2005). A. B. Klimov, C. Munos and J. L. Romero, arXiv:quant-ph/060511v1 (2005), Phys. Rev. A

\bibitem{ivanovich} I. D. Ivanovic, J. Phys. A, {\bf 14}, 3241 (1981).

\bibitem{ulf} U. Leonhardt, {\it Measuring the Quantum State of Light},  Cambridge University press, Cambridge (1997), p.19.

\bibitem{larry} L. S. Schulman, {\it Techniques and Applications of Path
Integation}, John Wiley and sons, Inc. (1981) p.38.



\bibitem{tal} S. Bandyopadhyay, P. O. Boykin, V. Roychowdhury and F. Vatan, arXiv:quant-ph/0103162v3 (2001), Algorithmica {\bf 34}, 512 (2002).

\bibitem{amir} A. Kalev, F.C. Khanna and M. Revzen, submitted for publication.

\bibitem{sam} S. Braunstein, A. Mann and M. Revzen, Phys. Rev. Lett. {\bf 68}, 3259
(1992).
\bibitem{epr} A. Einstein, B. Podolsky and N. Rosen, Phys. Rev {\bf 47}, 777 (1935).



\bibitem{schwinger} J. Schwinger, Proc. Nat. Acad. Sci. USA {\bf 46}, 560 (1960).

\bibitem{shim} A. Shimony in {\it The Dilemma of Einstein, Podolsky and Rosen - 60
years later,} Ann. Israel Phys. Soc. {\bf 12,} 163 (1996) Ed. by A. Mann and M.
Revzen.


%
\end{thebibliography}
\end{document}